\documentclass[aps,showpacs,twocolumn,superscriptaddress]{revtex4-2}
\usepackage[utf8]{inputenc}
\usepackage{amsfonts}
\usepackage{amssymb}
\usepackage{amsmath}
\usepackage{graphicx}
\usepackage{epsfig}
\usepackage{subfigure}
\usepackage{appendix}
\usepackage{hyperref}
\usepackage{color}

\hypersetup{hypertex=true, colorlinks=true, linkcolor=blue, urlcolor=blue, citecolor=blue}

\begin{document}

\title{Resonant dynamics of spin cluster in a periodically driven one-dimensional
Rydberg lattice}
\author{Jin-Qiu Xiong}
\affiliation{School of Physics and Optoelectronic Engineering, Foshan
University, Foshan, 528225, China}
\author{Yu-Hong Yan}
\affiliation{School of Physics and Optoelectronic Engineering, Foshan
University, Foshan, 528225, China}
\author{Xun-Da Jiang}
\affiliation{School of Physics and Optoelectronic Engineering, Foshan
University, Foshan, 528225, China}
\affiliation{Guangdong-HongKong-Macao Joint Laboratory for Intelligent Micro-Nano Optoelectronic Technology,
School of Physics and Optoelectronic Engineering, Foshan University, Foshan 528225, China}
\author{Yong-Yao Li}
\affiliation{School of Physics and Optoelectronic Engineering, Foshan
University, Foshan, 528225, China}
\affiliation{Guangdong-HongKong-Macao Joint Laboratory for Intelligent Micro-Nano Optoelectronic Technology,
School of Physics and Optoelectronic Engineering, Foshan University, Foshan 528225, China}
\author{Kun-Liang Zhang}
\email[Contact author: ]{zhangkl@fosu.edu.cn}
\affiliation{School of Physics and Optoelectronic Engineering, Foshan
University, Foshan, 528225, China} 
\affiliation{Guangdong-HongKong-Macao Joint Laboratory for Intelligent Micro-Nano Optoelectronic Technology,
School of Physics and Optoelectronic Engineering, Foshan University, Foshan 528225, China}
\date{\today}

\begin{abstract}
Rydberg lattice under facilitation conditions can feature kinetic constraints, leading to ballistic and nonergodic behavior at different detuning intensities. Here, we demonstrate that a resonant driving field can achieve effects similar to those under facilitation conditions. We focus on the relaxation dynamics of spin clusters in a periodically driven Rydberg spin lattice. Through an effective Hamiltonian for the domain walls of the spin cluster, it is shown that when the driving frequency is resonant with the Rydberg interaction, the spin cluster exhibits ballistic expansion with half the spreading rate compared to the case of facilitation conditions. However, near the resonant point, the spin cluster displays confinement behavior of the Bloch-like oscillations. These results demonstrate the rich dynamic behaviors in the driven Rydberg spin lattices and may find applications in quantum state manipulation. 
\end{abstract}

\maketitle

\section{Introduction}
\label{introduction}

Quantum systems subjected to external fields have garnered significant interest in recent years due to their potential to exhibit novel nonequilibrium phenomena and promising applications\cite{Rudner2020, else2020discrete, Collura2022, Longhi2012, Longhi2013, nag2014dynamical, luitz2017absence, agarwala2017effects, cao2022interaction, tiwari2024dynamical, bason2012high, Lin2014, zhang2022steady, mccullian2024coherent, li2017long, goldman2014periodically, zhang2024dynamics, zhang2024, zhang2024two}, such as Floquet topological phases \cite{Rudner2020}, discrete time crystals \cite{else2020discrete, Collura2022}, dynamical localization \cite{nag2014dynamical, luitz2017absence, agarwala2017effects, cao2022interaction, tiwari2024dynamical} and the control of quantum states through coherent driving \cite{bason2012high, Lin2014, zhang2022steady, mccullian2024coherent}.  Particularly, under the resonant condition where the frequency or strength of an applied field matches the energy spacing of the system, the interplay between the field and the system is enhanced, resulting in substantial changes in terms of long-range quantum correlations, quantum thermalization, and dynamic properties of the systems \cite{SanchisAlepuz2007, goldman2015periodically, bhakuni2020drive, zhang2020resonant, haldar2021dynamical, xu2022coherent, zhang2022steady, li2024collective, liu2025resonant, giudici2025fast}. This realm bridges fundamental questions about nonequilibrium quantum matters with applications in quantum simulation and quantum information.

The Rydberg lattice systems \cite{Zeiher2016, Labuhn2016, Marcuzzi2017, Bernien2017, Yang2019, Browaeys2020, Mukherjee2020, Li2020, Magoni2021,  Bluvstein2021,  Scholl2021, ebadi2021quantum, Mukherjee2022, Anand2024, Chen2024, Magoni2024, giudici2025fast, Pal2025a, Bombieri2025} are the ideal platforms for studying the  resonant dynamics of many-body quantum states, not only for the abilities for the fine-tuning of particle interactions and external fields, but also for the state of the art measurement techniques for the quantum states \cite{Browaeys2020}. The realm of Rydberg lattice systems with time-dependent driving is a rapidly developing frontier that combines the strong Rydberg interactions, programmable lattice geometries, and Floquet engineering to provide a powerful platform for exploring and manipulating quantum many-body dynamics \cite{Basak2018, guo2020optimized, zhao2023floquet, wu2025quantum}. Experimentally, the time-dependent driving fields can be readily implemented through the Rydberg lasers with time-varying Rabi frequency \cite{Bluvstein2021}. Recently, it was shown that the Rydberg lattices under facilitation conditions, where the laser detuning cancels out the nearest-neighbor interaction of atoms, feature rich dynamic behaviors \cite{Marcuzzi2017, Ostmann2019, Magoni2021, Magoni2022, Brady2025, Magoni2024} including Bloch oscillations of spin cluster \cite{Magoni2021}, coherent spin-phonon scattering \cite{Magoni2024} and nonclassical spin-phonon correlations \cite{Brady2025}. It turns out that the facilitation condition can be considered as a resonant condition for the spin cluster dynamics. Therefore, a resonant driving field may play a similar role in the facilitation condition in the aspect of spin cluster dynamics in the Rydberg lattice. The interplay between the driving field and Rydberg interaction may give rise to dynamics of spin cluster different from that under facilitation condition, and the modulated drive may provide another option or degree of freedom in parameter for manipulating the dynamics of spin clusters in the 1D Rydberg lattice.

In this work, we investigate the dynamics of spin cluster in a periodically driven one-dimensional (1D) Rydberg lattice. Specifically, we focus on the relaxation dynamics of spin cluster in this system under high-frequency driving field with the form $\mathit{\Omega} \left( t \right)=\mathit{\Omega}_{0}\cos\left(\omega t\right)$. We derive an effective Hamiltonian for the domain walls of the spin cluster, and show that there is a resonant frequency. 
When the driving frequency $\omega$ is resonant with the Rydberg interaction $V_{i,j}$, the spin clusters exhibit ballistic expansion with half the spreading rate compared to the case of facilitation conditions before the domain walls of spin cluster collide. After the first collision, the spin clusters exhibit subdiffusive dynamics. While for an initially Gaussian distribution with momentum $\pi$, the spin cluster exhibits ballistic expansion, and the total Rydberg density is conserved in the entire time domain. These results are demonstrated by the time evolution of Rydberg density and its variance related to the mean-square displacement. Moreover, we show that near the resonant driving frequency, the spin clusters manifest the  coexistence of oscillation and expansion dynamics, and the Bloch-like oscillation of the spin cluster, for different initial excitations. 

The rest of this paper is organized as follows: In Sec. \ref{model}, we introduce the Hamiltonian of the 1D Rydberg lattice with a periodic driving field, and discuss the dynamic properties in the spin cluster subspace under the condition of high frequency drive. In Sec. \ref{effective_H},  we derive the effective Hamiltonian for the domain walls of the spin cluster and obtain the resonant frequency.  In Sec. \ref{resonant}, we investigate the dynamics in the  resonant frequency and near the resonant frequency, and present numerical results of the evolutions of Rydberg density and its variance for different initial excitations. Finally, we summarize and discuss our findings in Sec. \ref{summary}, and conclude the results in Sec. \ref{conclusion}.

\section{Model and spin cluster subspace}

\label{model}

\begin{figure}[t]
\centering
\includegraphics[width=0.5\textwidth]{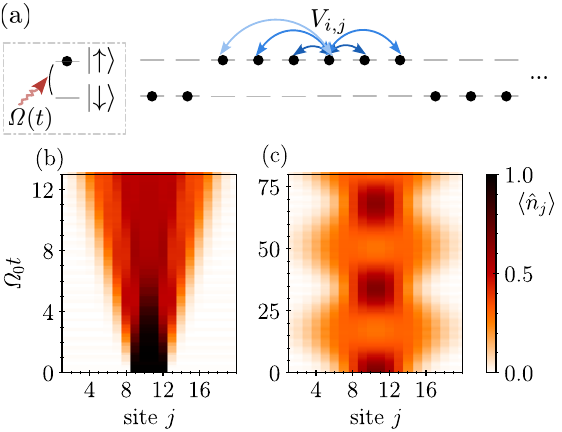}
\caption{(a) Schematic illustration of the 1D Rydberg lattice under periodic driving field  $\mathit{\Omega} \left( t \right)=\mathit{\Omega}_{0}\cos\left(\omega t\right)$. There are long-range interactions $V_{i,j}=\mathcal{V}_{0}/\left| i-j \right|^{6}$ between atoms in the Rydberg excited states $\left| \uparrow \right>$, and $\left| \downarrow \right>$ represents the ground state of an atom. The bottom panels present the numerical results of time evolutions of the Rydberg density $\left< \hat{n}_{j} \right>$ for different initial excitations and driving frequencies of driving fields. (a) Time evolutions of Rydberg density for simple spin cluster initial state under the resonant frequency $\omega=5.0867$, which is resonant with the Rydberg interaction and the system exhibits expansion dynamics of spin cluster. (b) Time evolutions of Rydberg density for a Gaussian distribution of spin cluster as initial state near the resonant frequency with $\omega=5.2867$. In contrast, the system exhibits oscillation dynamics in this case.  The size of the system is set as $N=20$, and the strength of the field and interaction are taken as $\mathit{\Omega}_{0}=1$ and $\mathcal{V}_{0}=5$, respectively.}
\label{fig_model}
\end{figure}

To investigate the resonant dynamics of spin clusters, we consider a 1D Rydberg lattice with $N$ sites under a periodic driving field, which is shown schematically in Fig. \ref{fig_model}(a). The Hamiltonian has the form 
\begin{equation}
\hat{\mathcal{H}}(t) =\sum_{i=1}^{N} \frac{\mathit{\Omega} \left( t \right)}{2} \hat{\sigma}%
_{i}^{x} +\sum_{i<j} V_{i,j}\hat{n}_{i} \hat{n}_{j},
\label{H_Ryd}
\end{equation}
where the operator $\hat{\sigma}_{i}^{x} =\left| \uparrow_{i} \right>\left< \downarrow_{i} \right|+\left| \downarrow_{i} \right> \left< \uparrow_{i} \right|$ couples the ground state $\left| \downarrow \right>$ and the Rydberg state $\left| \uparrow \right>$ of the $i$th  atom through the periodic driving field with the form $\mathit{\Omega} \left( t \right)=\mathit{\Omega}_{0}\cos\left(\omega t\right)$, the operator $\hat{n}_{i} =\left| \uparrow_{i} \right> \left< \uparrow_{i} \right|$ projects onto the Rydberg state for the $i$th  atom, and $V_{i,j}=\mathcal{V}_{0}/\left| i-j \right|^{6}$ is the long-range Van der Waals interaction between the Rydberg excited atoms at sites $i$ and $j$. Throughout this work, we assume $\mathit{\Omega}_{0}>0$ and $\mathcal{V}_{0}>0$, and periodic boundary conditions are applied. The dynamical controls for  nonequilibrium quantum many-body states of this model can be implemented experimentally in neutral ${}^{87}\textrm{Rb}$ atoms trapped in optical tweezers \cite{Bluvstein2021}. In this work, we are interested in the dynamics of the spin cluster under a periodic driving field. To this end, we take the spin cluster excitation as an initial state, and analyze this problem under the Hamiltonian of driven Rydberg lattice in Eq. (\ref{H_Ryd}).

The spin cluster states have the form
\begin{equation}
	\left| \psi \left( j_{1,}j_{2} \right) \right> =\prod_{j=j_{1}}^{j_{2}} \hat{\sigma}_{j}^{+} \prod_{i=1}^{N} \left| \downarrow_{i} \right>
	\label{state_j1j2}
\end{equation}
with $0<j_{1}\leqslant j_{2}<N$, where the raising operator of the $j$th atom is defined as $\hat{\sigma}_{j}^{+}=\left| \uparrow_{j} \right> \left< \downarrow_{j} \right|$.  Let us start with some heuristic derivations. First, note that in this state, the ground-state atoms that are far away from the  atoms inside the spin cluster, i.e., $l\ll j_{1}$ or $l\gg j_{2}$, are not subject to interactions $V_{ij}$. Thus before they are excited,  they can be regarded as a series of isolated two-level atoms in a global driving field $\mathit{\Omega} \left( t \right)/2$, and each of them can be described by the Hamiltonian $\hat{H}_{l}(t)=\mathit{\Omega} \left( t \right) \hat{\sigma}_{l}^{x}/2$. Their transition to the Rydberg excited states will be suppressed by the driving field, provided that the driving frequency $\omega$ is high enough, namely $\omega \gg \mathit{\Omega}_{0}$. In fact, for an atom at the $l$th site that is initially prepared on the ground state $\left| \downarrow_{l} \right>$, it can be checked that after time $t$, the probability of transition to Rydberg state $\left| \uparrow_{l} \right>$ is 
\begin{equation}
	P_{\mathrm{g}\rightarrow \mathrm{R}}(t)=\sin^{2}\left[\frac{\mathit{\Omega}_{0}\sin\left(\omega t \right)}{2\omega} \right],
	\label{PgR}
\end{equation}
which tends to zeros for high frequency drive $\omega \gg \mathit{\Omega}_{0}$.  In this case, only the atoms near the domain walls of spin cluster can be excited to Rydberg states to expand the cluster.

Second, the interaction between the atoms in Rydberg states can be replaced by the effective detuning $\mathit{\Delta}_{\mathrm{eff}}\sim \mathcal{V}_{0}$, under the mean-field approximation. Then, inside the spin cluster, each atom can be described by the Hamiltonian  $\hat{H}_{j}(t)=\mathit{\Omega} \left( t \right) \hat{\sigma}_{j}^{x}/2+\mathit{\Delta}_{\mathrm{eff}}\hat{n}_{j}$. According to the time-dependent perturbation theory, for an atom at site $j$ that is initially prepared in the Rydberg state $\left| \uparrow_{j} \right>$, the probability of transition to the ground state $\left| \downarrow_{j} \right>$ at  time $t$ is approximately 
\begin{equation}
	P_{\mathrm{R}\rightarrow \mathrm{g}}(t) \approx\frac{\mathit{\Omega}_{0}^{2} \sin^{2} \left[ \left( \mathit{\Delta}_{\mathrm{eff}} -\omega \right) t/2 \right]}{4\left( \mathit{\Delta}_{\mathrm{eff}} -\omega \right)^{2}},
	\label{PRg}
\end{equation}
which is suppressed when $\mathcal{V}_{0}\approx\omega \gg \mathit{\Omega}_{0}$ (see \ref{appendix_a} for the derivation). This achieves a similar effect to that under the facilitation conditions \cite{Marcuzzi2017, Ostmann2019, Magoni2021, Magoni2024, Brady2025}, where the detuning is set to cancel out the interaction between two adjacent atoms. In this parameter regime, it is interesting to further reveal the dynamic behavior of the domain walls of the spin cluster. 

In order to gain some intuitions and preliminary conclusions about the spin cluster dynamics, we perform numerical computations of time evolutions, observing the expansion dynamics of the spin cluster  when the driving frequency is taken as $\omega=5.0867$, and a slight deviation of the driving frequency leads to oscillation dynamics of the spin cluster, in Figs. \ref{fig_model}(b) and (c), respectively. In Fig. \ref{fig_model}(b), we present numerical  results of the Rydberg density $\left< \hat{n}_{j} \right>(t)=\left< \Psi \left( t \right) \right| \hat{n}_{j} \left| \Psi \left( t \right) \right>$ as a function of time for the initial excitation in Eq. (\ref{state_j1j2}), that is $\left| \Psi \left( 0 \right) \right>=\left| \psi \left( j_{1,}j_{2} \right) \right>$ with $j_{1}=9$ and $j_{2}=12$, for the first case, which is the simplest initial state to investigate the spin cluster dynamics and collision of domain walls of the driven Rydberg lattice. The evolved state 
\begin{equation}
\left| \Psi \left( t \right) \right> =\hat{\mathcal{T}} \exp \left[ -\mathrm{i} \int_{0}^{t} \hat{\mathcal{H}} \left( t^{\prime} \right) \textrm{d} t^{\prime} \right] \left| \Psi \left( 0 \right) \right>	
\end{equation}
is calculated under the time-dependent Hamiltonian in Eq. (\ref{H_Ryd}) using the fourth-order Runge-Kutta method with time steps of length $\Delta t=4\times 10^{-4}$ and the Hamiltonian is updated in each time step, where $\hat{\mathcal{T}}$ is the time-ordering operator and $\mathrm{i}=\sqrt{-1}$ (see \ref{appendix_b} for more details). 
Note that the spin cluster state in Eq. (\ref{state_j1j2}) consists of all momentum components. Therefore, it is worth considering Gaussian distribution with central momentum $k_{0}$ to investigate the contribution of different $k$ components to spin cluster dynamics.
In Fig. \ref{fig_model}(c), we take a Gaussian distribution with width $d=2$ as initial state, which has the following form
\begin{equation}
	\left| \Psi_{\mathrm{Gau.}} \left( 0 \right) \right>=\Lambda^{-1} \sum_{j_{1}} \mathrm{e}^{-\frac{\left( c-c_{0} \right)^{2}}{4d^{2}}}\mathrm{e}^{-\mathrm{i}k_{0}c}\left| \psi \left( j_{1},j_{1}+r_{0}-1 \right) \right>,
	\label{state_Gau}
\end{equation}
where $c=(j_{1}+j_{2})/2=j_{1}+(r_{0}-1)/2$ is the center of mass position of spin cluster (here $r_{0}=4$ is taken), $c_{0}=10.5$ and $k_{0}=\pi$ are, respectively, the average center of mass position and momentum, and $\Lambda$ is the normalization constant. We analyze the mechanism of these dynamics in the next section. 

\section{Effective Hamiltonian and resonant frequency}

\label{effective_H}

It is shown that under the condition of  high-frequency drive and strong interaction $\mathcal{V}_{0}\approx\omega \gg \mathit{\Omega}_{0}$, the probability of spin flipping far from the domain walls of the spin cluster  is suppressed. Moreover, we have seen that there exits a resonant driving frequency enabling the spin cluster to expand. In other words, in this parameter regime, the spin cluster can expand or shrink but cannot split. Thus we are able to project the Hamiltonian in Eq. (\ref{H_Ryd}) to the single cluster subspace $\left\{\left| \psi \left( j_{1,}j_{2} \right) \right>  \right\}$ by $\hat{H}_{\mathrm{eff}}=\hat{\mathcal{P}}\hat{\mathcal{H}}\hat{\mathcal{P}}^{-1}$, where the projector is defined as $\hat{\mathcal{P}}=\sum_{j_{1},j_{2}}\left| \psi \left( j_{1},j_{2} \right) \right> \left< \psi \left( j_{1},j_{2} \right) \right|$. Through this step, we obtain the spin cluster effective Hamiltonian 
\begin{equation}
	\begin{aligned}\hat{H}_{\mathrm{eff}} =&\mathit{\Omega}(t)\sum_{j_{1}<j_{2}} \left( \left| j_{1}+1,j_{2} \right\rangle +\left| j_{1},j_{2}+1 \right\rangle \right) \left\langle j_{1},j_{2} \right| +\textrm{H.c.} \\ &+\sum_{j_{1}<j_{2}} \sum_{l=1}^{j_{2}-j_{1}} \left( j_{2}-j_{1}-l+1 \right) V_{0,l}\left| j_{1},j_{2} \right\rangle \left\langle j_{1}, j_{2} \right|.
	\end{aligned}
	\label{H_eff}
\end{equation}
Here we denote the states defined in Eq. (\ref{state_j1j2}) as $\left| j_{1,}j_{2} \right>\equiv\left| \psi \left( j_{1,}j_{2} \right) \right>$ for brevity. This Hamiltonian, effectively describing the motion of the domain walls of the spin cluster, is equivalent to that for two hard-core bosons with time-dependent hopping and long-range interaction in a 1D lattice.

\begin{figure}[t]
\centering
\includegraphics[width=0.4\textwidth]{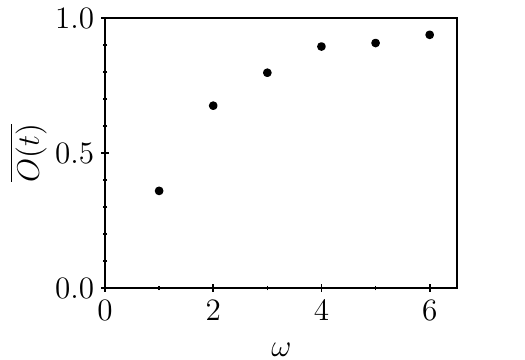}
\caption{Numerical results of time average overlap $\overline{O(t)}$ between the evolved states computed from the Rydberg spin Hamiltonian and the effective Hamiltonian for different driven frequency $\omega$. The initial states are all taken as the spin cluster states in Eq. (\ref{state_j1j2}) with the positions of domain walls $j_{1}=8$ and $j_{2}=11$. The size of the system is set as $N=18$, and the strength of the field and interaction are taken as $\mathit{\Omega}_{0}=1$ and $\mathcal{V}_{0}=5$, respectively. The results indicate that the effective Hamiltonian works well in the high frequency drive region.}
\label{fig_avg_overlap}
\end{figure}

Based on the above analyses, the effective Hamiltonian should works well in the high frequency drive region. To verify this point, we provide the comparison by the overlap between the evolved states computed from the Rydberg spin Hamiltonian and the effective Hamiltonian, which is defined as 
 \begin{equation}
 	O\left( t \right) =\left| \langle \Psi \left( t \right) |\Psi_{\mathrm{eff}} \left( t \right) \rangle \right|^{2},
 \end{equation}
where $\left| \Psi \left( t \right) \right> =\hat{\mathcal{T}} \exp \left[ -\mathrm{i} \int_{0}^{t} \hat{\mathcal{H}} \left( t^{\prime} \right) \textrm{d} t^{\prime} \right] \left| \Psi \left( 0 \right) \right>	$ and $\left| \Psi_{\mathrm{eff}} \left( t \right) \right> =\hat{\mathcal{T}} \exp \left[ -\mathrm{i} \int_{0}^{t} \hat{H}_{\mathrm{eff}} \left( t^{\prime} \right) \textrm{d} t^{\prime} \right] \left| \Psi \left( 0 \right) \right>	$, and the initial states are taken as the spin cluster states in Eq. (\ref{state_j1j2}). In Fig. \ref{fig_avg_overlap}, we present the numerical results of time average overlap 
\begin{equation}
	\overline{O(t)} =\frac{1}{t_{1}} \int_{0}^{t_{1}} O\left( t \right) \mathrm{d} t
\end{equation}
for different driven frequency $\omega$, and we take $t_{1}=10$. We can see that $\overline{O(t)}$ close to $1$ for high frequency drive $\omega \gg \mathit{\Omega}_{0}$, which verifies that the effective Hamiltonian works well in the high frequency drive region.

\begin{figure*}[t]
\centering
\includegraphics[width=1\textwidth]{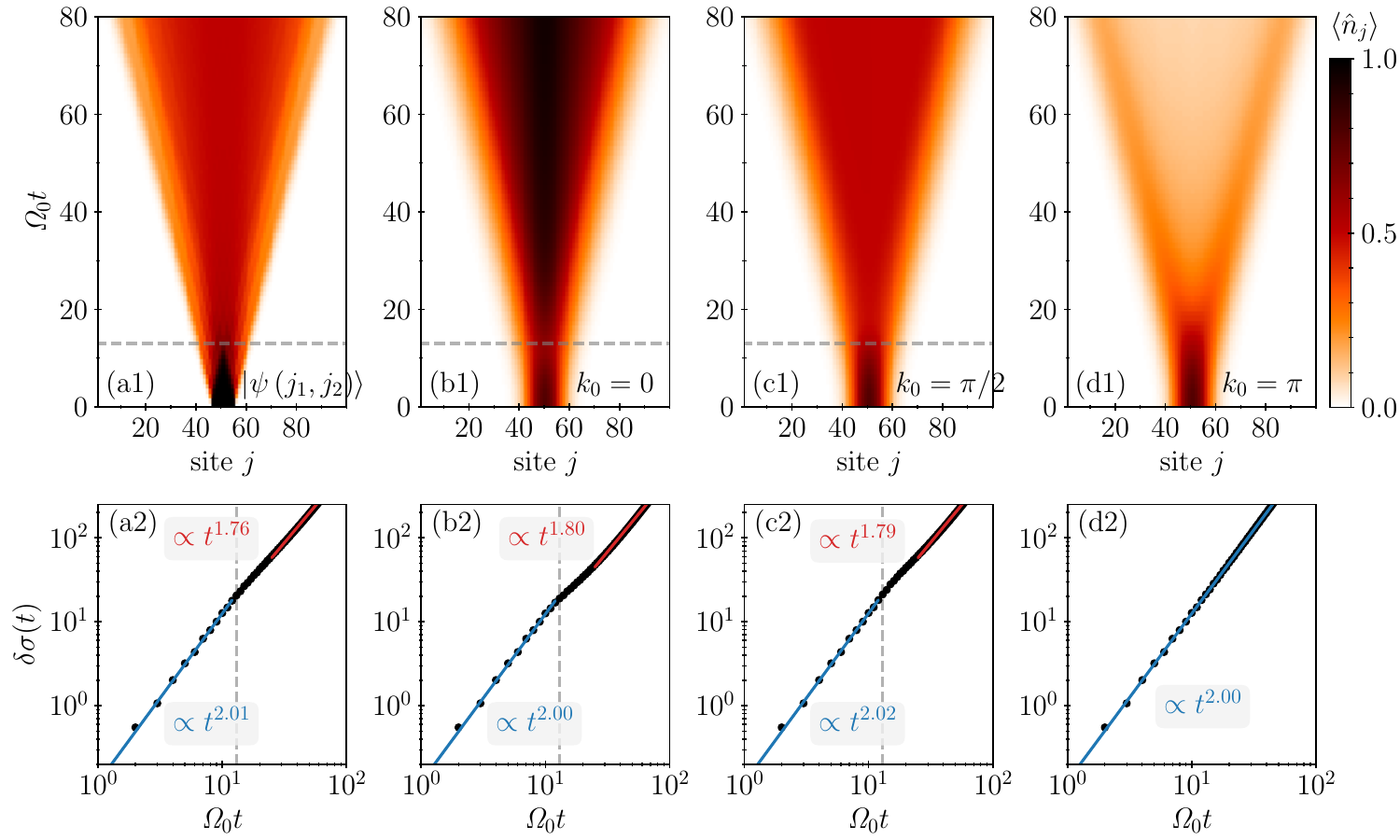}
\caption{Time evolutions of Rydberg density $\left< \hat{n}_{j} \right>$ and density variance $\delta \sigma(t)=\sigma(t)-\sigma(0)$ for the system under the resonant frequency of the driving field with different initial states. The frequency of the driving field is set as $\omega=5.0867$, which is resonant with the Rydberg interaction. The top panels present time evolution of Rydberg density $\left< \hat{n}_{j} \right>$ for the initial excitations of (a1) simple spin cluster $\left| \psi \left( j_{1},j_{2} \right) \right>$ in Eq. (\ref{state_j1j2}) with the positions of domain walls $j_{1}=46$ and $j_{2}=55$; the Gaussian distribution of the spin cluster in Eq. (\ref{state_Gau}) with different central momenta (b1) $k_{0}=0$, (c1) $k_{0}=\pi/2$ and (d1) $k_{0}=\pi$. The width, central position of the Gaussian distribution and size of the spin cluster are taken as $d=4$, $c_{0}=50.5$ and $r_{0}=10$, respectively. The bottom panels (a2)-(d2) show the density variance $\delta \sigma(t)$ on double-logarithmic scales (black dots) corresponding to the Rydberg density in (a1)-(d1), and the solid lines represent the linear fittings for the data. The size of the system is set as $N=100$, and the strength of the field and interaction are taken as $\mathit{\Omega}_{0}=1$ and $\mathcal{V}_{0}=5$, respectively. In the cases of  (a)-(c), it can be observed that the exponent $\beta$ of density variances decrease due to the collision of the domain walls. While the case in (c) with central momentum $k_{0}=\pi$, the spin cluster exhibits ballistic expansion with exponent $\beta=2$.}
\label{fig_resonant}
\end{figure*}

Transforming the basis of Hamiltonian (\ref{H_eff}) into the center of mass coordinate, i.e., $\left| j_{1},j_{2} \right> \rightarrow \left| c \right> \otimes \left| r \right>$ with $c=(j_{1}+j_{2})/2$ and $r=j_{2}-j_{1}+1$, and taking the Fourier transform for the center of mass coordinate  $\left| c \right> =1/\sqrt{2N} \sum_{k} \mathrm{e}^{\mathrm{i} kc} \left| k \right>$, the effective Hamiltonian can be written as the block diagonal form 
\begin{equation}
	\hat{H}_{\mathrm{eff}} =\sum_{k} \hat{H}_{k} \left( t \right) \otimes \left| k \right> \left< k \right|,
	\label{H_diagnal}
\end{equation}
with
\begin{equation}
	\begin{aligned}\hat{H}_{k}(t)=&2\mathit{\Omega} \left( t \right) \cos \left( \frac{k}{2} \right) \sum_{r=1}^{N-2} (|r+1\rangle \langle r|+\text{ H.c.})\\ &+\sum_{r=2}^{N-1}\mathcal{U} \left( r \right)|r\rangle \langle r|,\end{aligned}
	\label{H_eff_k}
\end{equation}
and $\mathcal{U} \left( r \right)= \sum_{l=1}^{r-1} (r-l)V_{0,l}$. The gradient of the potential is
\begin{equation}
	\mathcal{U} \left( r+1 \right)-\mathcal{U} \left( r \right)\approx \sum_{l=1}^{r-1} \frac{\mathcal{V}_{0}}{l^{6}} \equiv \mathcal{F},
\end{equation}
which is a constant for large $r$. This means that for the fixed $k$ and $t$, the Hamiltonian $\hat{H}_{k}(t) $ describes a hopping particle in a lattice with skew potential $\mathcal{F}r$, which is closely related to the Bloch oscillations \cite{Bloch1929, Wannier1960}. 

In the large-$N$ limit and for a finite $\mathcal{F}$, the eigenstates of the Hamiltonian in Eq. (\ref{H_eff_k}) are all localized, forming the celebrated  Wannier-Stark  ladder states \cite{Wannier1960}. We consider the eigenstates localized at $r\gg 1$, which have the form 
\begin{equation}
	\left| \phi_{n,k} \right\rangle =\sum_{r} J_{r-n}(z_k)\left| r \right\rangle,
	\label{WS_states}
\end{equation}
with energies $E_{n,k}=n\mathcal{F}$ and positive integer $n\gg 1$. Here $J_{r-n}(z_k)$ is the Bessel function of the first kind, and $z_k=-4\mathit{\Omega} \left( t \right) \cos \left( k/2 \right)/\mathcal{F}$. Obviously, Eq. (\ref{WS_states}) characterizes a series of eigenstates localized near site $n$, and the localization length decreases as $|z_{k}|$ decreases. Additionally, we note that while the eigenstates are dependent on time $t$, the energy levels $E_{n,k}$ are time-independent, which is crucial for the subsequent analysis. 

Due to the block decomposition of the effective Hamiltonian in Eq. (\ref{H_diagnal}), it is possible to consider the dynamics in each subspace of $k$ individually.  Considering the time evolution of an initial state
 \begin{equation}
 	\left| \Psi \left( 0 \right) \right> =\sum_{k} \mathcal{A}_{k} \left| \Psi_{k} \left( 0 \right) \right> \otimes \left| k \right>,
 \end{equation}
since the energy levels are time-independent, under the basis $\left| \phi_{n,k} \right\rangle $, the time evolution of $\left| \Psi_{k} \left( 0 \right) \right>$ assumes the form
\begin{equation}
\begin{aligned}
|\Psi_{k} (t)\rangle =&\sum_{n} \mathcal{C}_{n,k}(t)\mathrm{e}^{-\mathrm{i} \int_{0}^{t} E_{n,k}\mathrm{d} t^{\prime}} \left| \phi_{n,k} \right>\\ =&\sum_{n} \mathcal{C}_{n,k}(t)\mathrm{e}^{-\mathrm{i} n\mathcal{F} t} \left| \phi_{n,k} \right>.	
\end{aligned}
\label{Psi_kt}
\end{equation}
Submitting Eq.(\ref{Psi_kt}) into the Schr\"{o}dinger equation 
\begin{equation}
	\mathrm{i} \frac{\partial}{\partial t} |\Psi_{k} (t)\rangle =\hat{H}_{k} \left( t \right) |\Psi_{k} (t)\rangle,
\end{equation}
and left-hand side multiplied by $\left< \phi_{m,k} \right|$, we obtain the equation of motion for amplitude $\mathcal{C}_{n,k}(t)$, that is 
\begin{equation}
	\frac{\partial \mathcal{C}_{m,k}(t)}{\partial t} =-\sum_{n} \mathcal{C}_{n,k}(t) \mathrm{e}^{\mathrm{i} (m-n)\mathcal{F} t} \left\langle \phi_{m,k} \right| \frac{\partial}{\partial t} \left| \phi_{n,k} \right\rangle .
\end{equation}
Note, however, that $\left| \phi_{n,k} \right\rangle$ is time dependent. Submitting the expression of $\left| \phi_{n,k} \right\rangle$ [see Eq. (\ref{WS_states})] into the above equation, and using the recurrence relation of Bessel function $2 \partial_{x} J_{n}(x)=J_{n-1}(x)-J_{n+1}(x)$, we have
\begin{widetext}
\begin{equation}
	\frac{\partial \mathcal{C}_{m,k}(t)}{\partial t} =-\frac{2\mathit{\Omega}_{0} \omega}{\mathcal{F}} \cos \left( \frac{k}{2} \right) \sin (\omega t)\left[\mathcal{C}_{m-1,k}(t) \mathrm{e}^{\mathrm{i} \mathcal{F} t} -\mathcal{C}_{m+1,k}(t) \mathrm{e}^{-\mathrm{i} \mathcal{F} t} \right].
	\label{Cmt1}
\end{equation}
In the regime of high frequency drive $\mathcal{V}_{0}\approx\omega \gg \mathit{\Omega}_{0}$, the condition $\left| \omega -\mathcal{F} \right| \ll \mathcal{F}$ is satisfied since $\mathcal{F}\approx \mathcal{V}_{0}$. Then we are able to take the rotating-wave approximation, and Eq. (\ref{Cmt1}) becomes
\begin{equation}
	\mathrm{i} \frac{\partial \mathcal{C}_{m,k}(t)}{\partial t} =\frac{\mathit{\Omega}_{0} \omega}{\mathcal{F}} \cos \left( \frac{k}{2} \right) \left[ \mathcal{C}_{m-1,k}(t) \mathrm{e}^{\mathrm{i} \left( \mathcal{F} -\omega \right) t} +\mathcal{C}_{m+1,k}(t) \mathrm{e}^{-\mathrm{i} \left( \mathcal{F} -\omega \right) t} \right] .
	\label{Cmt2}
\end{equation}
\end{widetext}

Noting that in the resonant driving frequency $\omega = \mathcal{F}$, the above equation reduces to
\begin{equation}
	\mathrm{i} \frac{\partial \mathcal{C}_{m,k}(t)}{\partial t} =\mathit{\Omega}_{0} \cos \left( \frac{k}{2} \right) \left[ \mathcal{C}_{m-1,k}(t)+\mathcal{C}_{m+1,k}(t) \right] ,
	\label{Cmt3}
\end{equation}
which is nothing but the Schr\"{o}dinger equation for a moving particle in a uniform tight-binding chain with a $k$-dependent hopping amplitude  $\mathit{\Omega}_{0}\cos \left( k/2 \right)$. In the evolved state in Eq. (\ref{Psi_kt}), basis $\left| \phi_{n,k} \right\rangle $ describes a state localized near $n$, which characterizes the size of a spin cluster with amplitude $\mathcal{C}_{n,k}(t)$. Thus in this case, the spin cluster would experience ballistic expansion, despite the presence of the long-range interaction.

\begin{figure}[h]
\centering
\includegraphics[width=0.5\textwidth]{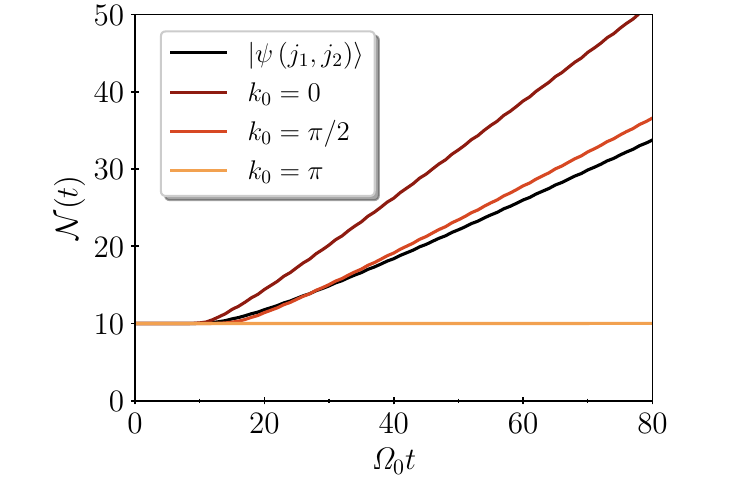}
\caption{Total Rydberg density $\mathcal{N}(t)$ as a function of time for the resonant cases in Fig. \ref{fig_resonant}(a1)-(d1). The frequency of the driving field is set as the resonant value $\omega=5.0867$. The size of the system is set as $N=100$, and the strength of the field and interaction are taken as $\mathit{\Omega}_{0}=1$ and $\mathcal{V}_{0}=5$, respectively. Combining with the results in Fig. \ref{fig_resonant}, one can see that the total Rydberg density increase due to the collision of domain walls, and the total Rydberg density is conserved for the case with central momentum $k_{0}=\pi$.}
\label{fig_resonant_Nt}
\end{figure}

\begin{figure*}[tbh]
\centering
\includegraphics[width=1\textwidth]{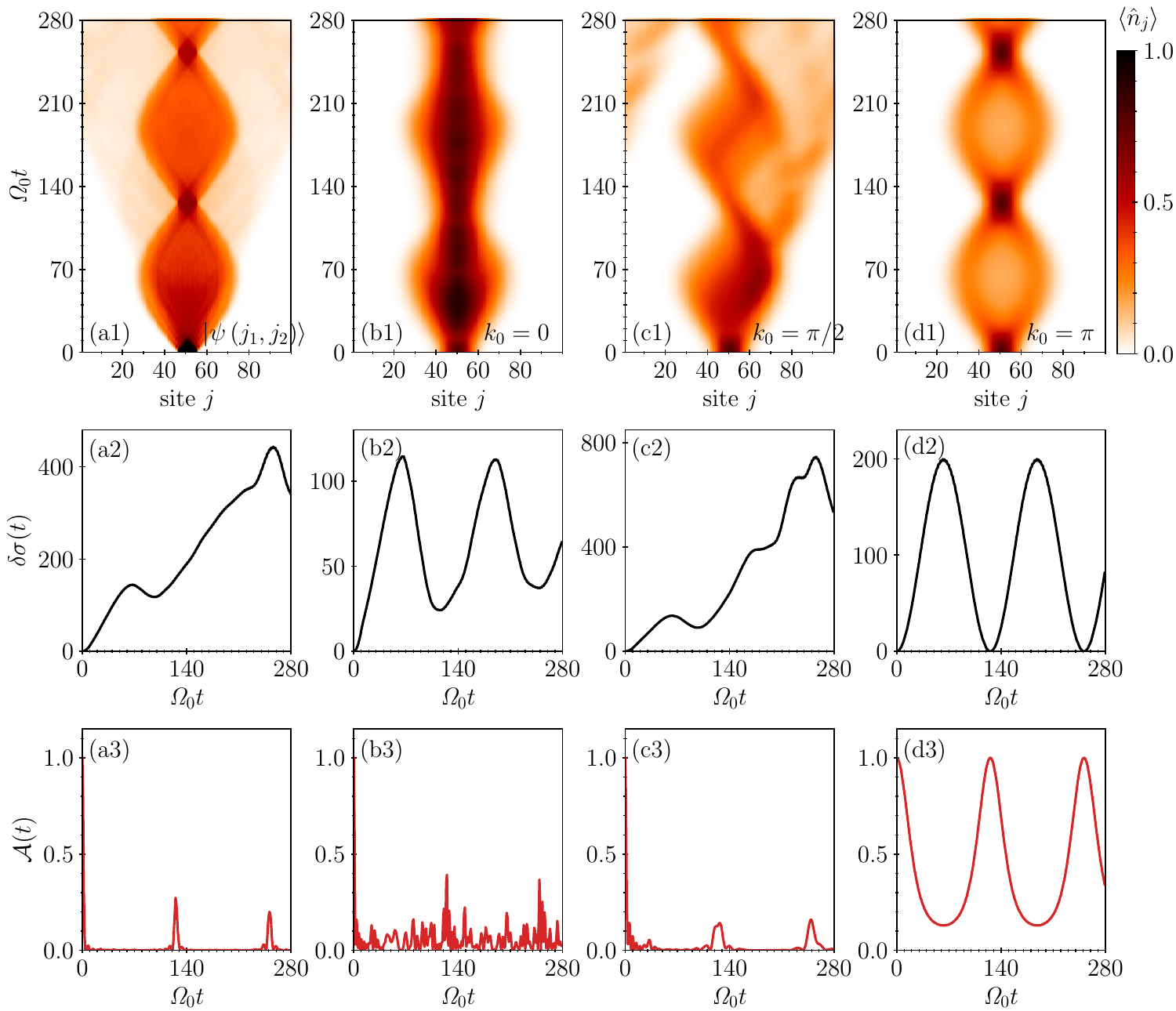}
\caption{Time evolution of Rydberg density $\left< \hat{n}_{j} \right>$, density variance $\delta \sigma(t)=\sigma(t)-\sigma(0)$ and autocorrelation function $\mathcal{A}(t)$ for the system near the resonant frequency of the driving field with different initial states. The frequency of the driving field is set as $\omega=5.1367$. The top panels (a1)-(d1) present time evolutions of Rydberg density $\left< \hat{n}_{j} \right>$ for the initial excitations taken as the same as those in Fig. \ref{fig_resonant}(a1)-(d1). The middle panels (a2)-(d2) show the density variance $\delta \sigma(t)$ corresponding to the Rydberg density in (a1)-(d1). The bottom panels show the autocorrelation function $\mathcal{A}(t)$ between the initial states and evolved states corresponding to the cases in (a1)-(d1).  The size of the system is set as $N=100$, and the strength of the field and interaction are taken as $\mathit{\Omega}_{0}=1$ and $\mathcal{V}_{0}=5$, respectively. The periodic boundary condition is taken for the system. For the cases in (a)-(c), the systems exhibit confinement dynamics of spin cluster without revival due to the near resonant condition and the collision of spin cluster domain walls, while for the Gaussian distribution with $k_{0}=\pi$ in (d), the spin cluster exhibits oscillation dynamics with high-amplitude revival due to the absent of the collision.}
\label{fig_oscillation}
\end{figure*}

\section{Resonance and near resonance dynamics of spin cluster}
\label{resonant}
\subsection{Resonance dynamics}

To verify the above analysis and provide further conclusions, we perform numerical simulations of time evolutions under the effective Hamiltonian in Eq. (\ref{H_eff}). This allows us to handle a system with a larger $N$, and the results of spin cluster dynamics are expected to accord with that obtained from  the Hamiltonian in Eq. (\ref{H_Ryd}) under the condition of  high-frequency drive $\mathcal{V}_{0}\approx\omega \gg \mathit{\Omega}_{0}$.

In Fig. \ref{fig_resonant}(a1), we present the result of time evolution of Rydberg density $\left< \hat{n}_{j} \right> (t)$ for the spin cluster that is initially 
prepared in a state of the form in Eq. (\ref{state_j1j2}) with fixed positions of domain walls. Here we take the interaction strength as $\mathcal{V}_{0}=5$, and then the resonant frequency is $\mathcal{F}=\sum_{l=1}^{r-1} \mathcal{V}_{0}/l^{6}\approx 5.0867$. The time evolution is numerically  computed using the fourth-order Runge-Kutta method with time steps of length $\Delta t=2\times 10^{-4}$. At first glance, the profile of the spin cluster experiences ballistic expansion, while there is a change in Rydberg density at time $t\approx 13 \mathit{\Omega}_{0}$, which is marked by a gray dashed line.

In order to gain a deeper understanding of the evolution of Rydberg density, we introduce the Rydberg density variance $\sigma (t)$ \cite{Magoni2024}, which is defined as
\begin{equation}
	\sigma(t)=\sum_{j=1}^N j^2 \frac{\left\langle \hat{n}_j\right\rangle(t)}{\mathcal{N}(t)}-\left(\sum_{j=1}^N j \frac{\left\langle \hat{n}_j\right\rangle(t)}{\mathcal{N}(t)}\right)^2,
\end{equation}
where 
\begin{equation}
	\mathcal{N}(t)=\sum_{l=1}^N\left\langle \hat{n}_{l}\right\rangle (t)
\end{equation}
denotes the total Rydberg density. The density variance $\sigma (t)$ measures the spreading dynamics of spin cluster and relates to the mean square displacement \cite{steinigeweg2017real, Lev2017, lm2022logarithmic, sierant2023slow, zhang2024two}, which is expected to increase over time as the form 
\begin{equation}
	\delta \sigma(t)=\sigma(t)-\sigma(0) \sim t^\beta ,
\end{equation}
where $\sigma(0)$ is the density variance of the spin cluster at $t=0$. It is expected to observe $\beta=2$ for the ballistic transport. In Fig. \ref{fig_resonant}(a2) and Fig. \ref{fig_resonant_Nt}, we present the Rydberg density variance as a function of time on double-logarithmic scales and the total Rydberg density as a function of time, respectively, corresponding to the Rydberg density in Fig. \ref{fig_resonant}(a1). We can see that before $t\approx 13 \mathit{\Omega}_{0}$, the spin cluster did show ballistic expansion with $\beta =2.01$, while for a larger time $t\gtrsim 25 \mathit{\Omega}_{0}$ the factor drops to $\beta =1.76$. Moreover, in Fig. \ref{fig_resonant_Nt},  it is shown that the total Rydberg density is conserved before $t\approx 13 \mathit{\Omega}_{0}$ and grows linearly for larger time. These behaviors arise because the number of spin clusters is conserved throughout the high-frequency drive process, and the size of the initial spin cluster $r_0$ is finite. Therefore when the size of the spin cluster shrinks to $r=1$, the effect of an effective hard-core repulsion potential appears. This is similar to the situation of the Rydberg lattice under the facilitation condition but without the driving field \cite{Marcuzzi2017, Magoni2024}. Note that the hopping amplitude in Eq. (\ref{Cmt3}) is half of the static case with $\omega=0$ for the hopping amplitude in Eq. (\ref{H_eff_k}). This leads to a time delay in the collision of the spin cluster's domain walls.

It is also worth noting that the hopping amplitude vanishes for the momentum component $k=\pi$, which would prevent any expansion or shrinkage of the spin cluster, leading to the ballistic expansion and the conservation of total Rydberg density of the spin cluster. In Figs. \ref{fig_resonant}(b1)-(d1), Figs. \ref{fig_resonant}(b2)-(d2) and Fig. \ref{fig_resonant_Nt}, we present the numerical results of Rydberg density $\left< \hat{n}_{j} \right>(t)$, Rydberg density variance $\delta \sigma(t)$ and total Rydberg density $\mathcal{N}(t)$, respectively, for the initial Gaussian distribution in Eq. (\ref{state_Gau}) with width $d=4$ for different momentum $k_{0}$. The numerical results show that the cases with $k_{0}=0$ and $\pi/2$ are similar to that of the initial state $\left| \psi \left( j_{1,}j_{2} \right) \right>$: the spin clusters first experience ballistic expansion and then exhibit subdiffusive dynamics at a larger time. However, the results in Figs. \ref{fig_resonant}(d1), (d2) and Fig. \ref{fig_resonant_Nt} indicate that for case with momentum $k_{0}=\pi$,    the spin cluster indeed exhibit ballistic expansion, and the total Rydberg density is conserved at all times. The primary interest in this effect stems from its occurrence in a system featuring the combined effects of long-range interactions and time-dependent driving.

\begin{figure}[t]
\centering
\includegraphics[width=0.5\textwidth]{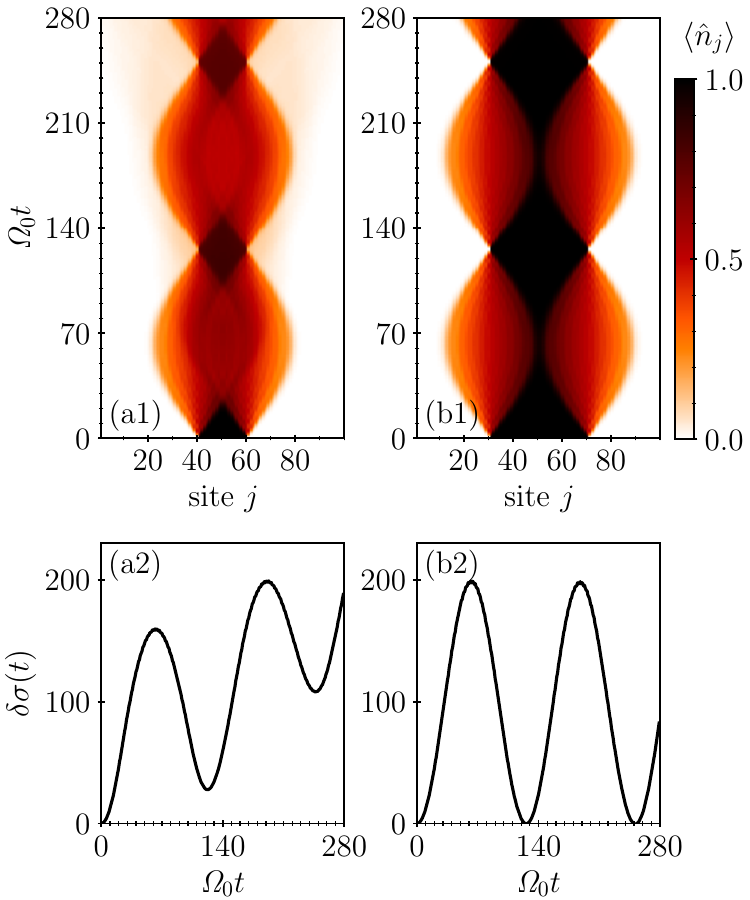}
\caption{Time evolution of Rydberg density $\left< \hat{n}_{j} \right>$, density variance $\delta \sigma(t)=\sigma(t)-\sigma(0)$ for the system near the resonant frequency of the driving field with spin cluster initial states $\left| \psi \left( j_{1},j_{2} \right) \right>$. The frequency of the driving field is set as $\omega=5.1367$. The top panels (a1) and (b1) present time evolutions of Rydberg density $\left< \hat{n}_{j} \right>$ for the initial excitations with $(j_{1},j_{2})=(41,60)$ and $(j_{1},j_{2})=(31,70)$, respectively. The bottom panels (a2) and (b2) show the density variance $\delta \sigma(t)$ corresponding to the Rydberg density in (a1) and (b1). The size of the system is set as $N=100$, the strength of the field and interaction are taken as $\mathit{\Omega}_{0}=1$ and $\mathcal{V}_{0}=5$, respectively, and the periodic boundary condition is taken for the system. These indicate that the domain walls of spin cluster can no longer collide when the initial spin cluster size $r_{0}=j_{2}-j_{1}+1$ is larger than the amplitude of oscillation.}
\label{fig_oscillation_2}
\end{figure}

\subsection{Dynamics near the resonant point}

Now we investigate the effect of a minor deviation on the resonant frequency. Setting $\omega=\mathcal{F}+\delta \omega$ with $\delta \omega \ll \mathcal{F}$, Eq. (\ref{Cmt2}) becomes
\begin{equation}
	\mathrm{i} \frac{\partial \mathcal{C}_{m,k}(t)}{\partial t} =\mathit{\Omega}_{k} \left[ \mathcal{C}_{m-1,k}(t)\mathrm{e}^{-\mathrm{i}\delta \omega}+\mathcal{C}_{m+1,k}(t) \mathrm{e}^{\mathrm{i}\delta \omega}\right] ,
\end{equation}
where $\mathit{\Omega}_{k}=\mathit{\Omega}_{0} \omega \cos \left( k/2 \right) /\mathcal{F}\approx \mathit{\Omega}_{0} \cos \left( k/2 \right)$. The above equation indicates that the spin cluster may exhibits periodic dynamics like Bloch oscillation with period $T=2\pi \mathit{\Omega}_{0}/\delta \omega$ for certain initial excitation.

In Figs. \ref{fig_oscillation}(a1)-(d1) and Figs. \ref{fig_oscillation}(a2)-(d2), we present the numerical results of Rydberg density $\left< \hat{n}_{j} \right>(t)$ and Rydberg density variance $\delta \sigma(t)$, respectively, for the same initial states as that in Fig. \ref{fig_resonant}, while the driving frequency is taken as $\omega=5.1367$ with $\delta \omega=0.05$.  In addition, in order to characterize the amplitude revival of the evolved states,  we introduce the autocorrelation function
\begin{equation}
	\mathcal{A} (t)=\left| \left< \psi (0) | \psi (t) \right> \right|^{2},
\end{equation}
with the corresponding results shown in Figs. \ref{fig_oscillation}(a3)-(d3). 

The results in Figs. \ref{fig_oscillation}(a1)-(a3) and Figs. \ref{fig_oscillation}(c1)-(c3) demonstrate that for both initial states of $\left| \psi \left( j_{1,}j_{2} \right) \right>$ and Gaussian distribution with $k_{0}=\pi/2$, the systems exhibits the coexistence of oscillation and expansion dynamics of spin clusters, while for the Gaussian distribution with $k_{0}=0$ in Figs. \ref{fig_oscillation}(b1)-(b3), the systems exhibit confinement dynamics of spin cluster without revival. These phenomena are mainly attributed to the collision of spin cluster domain walls. Notably, for the Gaussian distribution with $k_{0}=\pi$ in Figs. \ref{fig_oscillation}(d1)-(d3), the spin cluster exhibits oscillation dynamics with high amplitude revival, and as expected, the period is $T=2\pi\mathit{\Omega}_{0}/\delta \omega \approx 125.66\mathit{\Omega}_{0}$.

Moreover, the initial spin cluster size $r_0$ is closely related to the collision of domain walls of spin cluster. For the resonant case, the collision time of domain walls  is proportional to $r_0$,  while for oscillation dynamics near the resonant frequency, the domain walls can no longer collide when $r_0$ is larger than the amplitude of oscillation. In this case, the spin cluster does not expand and only oscillation dynamics exists. 
In Fig. \ref{fig_oscillation_2}, we present the numerical results for time evolution of Rydberg density and density variance near the resonant frequency for $r_{0}=20$ and $r_{0}=40$. In comparison with the cases with $r_{0}=10$ in Fig. \ref{fig_oscillation}(a1) and $r_{0}=20$ in Fig. \ref{fig_oscillation_2}(a1), we can see that for the case with $r_{0}=40$, the two domain walls move independently in Figs. \ref{fig_oscillation_2}(b1) and (b2), and the oscillation amplitude of density variance $\delta \sigma(t)$ does not increase as time.

\section{Summary and Discussion}
\label{summary}

In summary, we investigate the relaxation dynamics of the spin cluster in a 1D Rydberg lattice system under high-frequency driving field, and find a resonant frequency where the spin clusters exhibit ballistic expansion with half the spreading rate compared to the case under facilitation conditions before the domain walls of the spin cluster collide. After the first collision, the spin clusters exhibit subdiffusive dynamics, while for an initial excitation with a Gaussian distribution of momentum $\pi$, the spin cluster exhibits ballistic expansion, and the total Rydberg density is conserved in the entire time domain. Moreover, we show that near the resonant driving frequency, the spin clusters manifest the  coexistence of oscillation and expansion dynamics, and the Bloch-like oscillation of the spin cluster, for different initial excitations. 

These features are highly promising for experimental verification in future research. In the Rydberg lattice system, the time-dependent driving fields can be readily implemented experimentally through the Rydberg lasers with time-varying Rabi frequency \cite{Bluvstein2021}. The single-site addressability allows for the individual excitation of atoms from the ground to the Rydberg state \cite{Bernien2017, de2017optical}, and thus the spin cluster state is feasible in experiments. While the preparation of initial state with Gaussian distribution is more difficult, the former work \cite{surace2021scattering} suggests a protocol to generating a narrow momentum distribution of spin wave packet. For example, on the left side of the spin chain, a spatially localized spin flip is generated by spatially inhomogeneous fields. The sharp variation in the field parameters allows only components with specific momenta to propagate to the right, thereby achieving momentum filtering. It is also possible to implement a similar protocol to our work by introducing an inhomogeneity into the lattice.

\section{Conclusion}
\label{conclusion}

The modulated drive in our work does not require modulating the detuning precisely to control the dynamics, and may provide another option or degree of freedom in parameter for manipulating the dynamics of spin clusters in the 1D Rydberg lattice. 
Moreover, we find that there is a time delay in the collision of the spin cluster’s domain walls in comparison to the case with the facilitation condition because  the effective hopping amplitude is half of that of the static case. it is noteworthy that ballistic transport of spin clusters occurs in a system with long-range interactions and time-dependent driving. These results demonstrate the rich dynamic behaviors in the driven Rydberg spin lattices.

Our conclusions are applicable to the 1D Rydberg lattice with other types of long-range interactions, such as the dipole-dipole interaction with the form  $V_{i,j}=\mathcal{V}_{0}/\left| i-j \right|^{3}$. The resonant dynamics of the spin cluster can also be observed in the transverse Ising chain in the presence of a
longitudinal field \cite{simon2011quantum, cai2011, kormos2017real, verdel2020, lerose2020quasilocalized, verdel2023, lagnese2024detecting}. In this case, the longitudinal field can be periodically modulated to be resonant with the transverse field. In the future, it would be interesting to study the impacts of spin-phonon interactions \cite{Magoni2024, Brady2025} on the resonant dynamics in the driven Rydberg lattices.

\acknowledgments This work was supported by the National Natural Science Foundation of China (Grants No. 12505015, No. 12305013, and No. 12274077),  the Natural Science Foundation of Guangdong Province (Grants No. 2024A1515110222, and  No. 2023A1515010770),  and the Research Fund of Guangdong-HongKong-Macao Joint Laboratory for Intelligent Micro-Nano Optoelectronic Technology (No. 2020B1212030010).

\label{A} \setcounter{equation}{0} \renewcommand{\theequation}{A%
\arabic{equation}}
\label{A} \setcounter{figure}{0} \renewcommand{\thefigure}{A%
\arabic{figure}}

\setcounter{section}{0} \renewcommand{\thesection}{APPENDIX A}
\section{The transition probabilities}
\label{appendix_a}

In this Appendix, we present the derivations of the transition probabilities in Eq. (\ref{PgR}) and Eq. (\ref{PRg}), respectively.

For the Hamiltonian $\hat{H}_{l}(t)=\mathit{\Omega}_{0}\cos\left(\omega t\right) \hat{\sigma}_{l}^{x}/2$, the transition probability can be solved exactly. The time evolution for the initial state $\left| \varphi (0) \right\rangle=\left| \downarrow \right\rangle$ reads
\begin{equation}
	\left| \varphi (t) \right\rangle =\sum_{n=\pm} C_{n}(t)\mathrm{e}^{-\mathrm{i} \int_{0}^{t} E_{n}\left( t' \right) \mathrm{d} t^{\prime}} \left| n \right\rangle,
\end{equation}
where the instantaneous eigenenergies of $\hat{H}_{l}(t)$ are $E_{\pm}\left( t \right)=\pm \mathit{\Omega}_{0}\cos\left(\omega t\right)/2$, with time-independent eigenstates $\left| \pm \right\rangle=(\left| \uparrow \right\rangle\pm \left| \downarrow \right\rangle)/\sqrt{2}$, and the initial condition is $C_{\pm}=\pm 1/\sqrt{2}$. The coefficient $C_{n}(t)$ is time-independent, since
\begin{equation}
	\frac{\partial C_{n}(t)}{\partial t} =-\sum_{m=\pm} C_{m}(t)\mathrm{e}^{\mathrm{i} \theta_{m,n} \left( t \right)} \left\langle m \right| \frac{\partial}{\partial t} \left| n \right\rangle =0,
\end{equation}
where $\theta_{m,n} \left( t \right) =\int_{0}^{t} \left[ E_{n}\left( t^{\prime} \right) -E_{m}\left( t^{\prime} \right) \right] \mathrm{d} t^{\prime}.$ Then the evolved state becomes 
\begin{equation}
	\begin{aligned}\left| \varphi (t) \right\rangle &=\sum_{n=\pm} \frac{n}{\sqrt{2}} \exp \left[ -\mathrm{i} n\frac{\mathit{\Omega}_{0} \sin \left( \omega t \right)}{2\omega} \right] \left| n \right\rangle\\ &=-\mathrm{i} \sin \left[ \frac{\mathit{\Omega}_{0} \sin \left( \omega t \right)}{2\omega} \right] \left| \uparrow \right\rangle +\cos \left[ \frac{\mathit{\Omega}_{0} \sin \left( \omega t \right)}{2\omega} \right] \left| \downarrow \right\rangle.\end{aligned}
\end{equation}
That is, the transition probability to Rydberg state is 
\begin{equation}
	P_{\mathrm{g}\rightarrow \mathrm{R}}(t)=\sin^{2}\left[\frac{\mathit{\Omega}_{0}\sin\left(\omega t \right)}{2\omega} \right].
\end{equation}

However, for Hamiltonian  $\hat{H}_{j}(t)=\mathit{\Omega}_{0}\cos\left(\omega t\right) \hat{\sigma}_{j}^{x}/2+\mathit{\Delta}_{\mathrm{eff}}\hat{n}_{j}$, the transition probability can be evaluated by the time-dependent perturbation theory \cite{sakurai2020modern}. Under the condition $\mathit{\Omega}_{0}/2 \ll \mathit{\Delta}_{\mathrm{eff}}$, the first term of $\hat{H}_{j}(t)$ can be treated as a perturbation. Up to first order, the amplitude of transition from the Rydberg state to ground state is 
\begin{equation}
\begin{aligned}C_{\mathrm{R} \rightarrow \mathrm{g}}\left( t \right)&\approx -\mathrm{i} \int_{0}^{t} \frac{\mathit{\Omega}_{0} \cos \left( \omega t^{\prime} \right)}{2} \mathrm{e}^{-\mathrm{i} \mathit{\Delta}_{\mathrm{eff}} t^{\prime}} \mathrm{d} t^{\prime}\\ &=\frac{\mathit{\Omega}_{0}}{4} \left[ \frac{\mathrm{e}^{-\mathrm{i} (\mathit{\Delta}_{\mathrm{eff}} -\omega )t} -1}{\mathit{\Delta}_{\mathrm{eff}} -\omega} +\frac{\mathrm{e}^{-\mathrm{i} (\mathit{\Delta}_{\mathrm{eff}} +\omega )t} -1}{\mathit{\Delta}_{\mathrm{eff}} +\omega} \right] .\end{aligned}
\end{equation}
This can be simplified through the rotating-wave approximation, that is, when $\mathit{\Delta}_{\mathrm{eff}} +\omega \gg \left| \mathit{\Delta}_{\mathrm{eff}} -\omega \right|$, the second term in the above equation can be omitted. Then we have 
\begin{equation}
	\begin{aligned}C_{\mathrm{R} \rightarrow \mathrm{g}}\left( t \right) &\approx \frac{\mathit{\Omega}_{0}}{4} \frac{\mathrm{e}^{-\mathrm{i} (\mathit{\Delta}_{\mathrm{eff}} -\omega )t} -1}{\mathit{\Delta}_{\mathrm{eff}} -\omega}\\ &=-\mathrm{i} \mathrm{e}^{-\mathrm{i} (\mathit{\Delta}_{\mathrm{eff}} -\omega )t/2} \frac{\mathit{\Omega}_{0} \sin \left[ (\mathit{\Delta}_{\mathrm{eff}} -\omega )t/2 \right]}{2\left( \mathit{\Delta}_{\mathrm{eff}} -\omega \right)},\end{aligned}
\end{equation}
and the transition probability from the Rydberg state to ground state is 
\begin{equation}
	P_{\mathrm{R} \rightarrow \mathrm{g}}(t)=\left| C_{\mathrm{R} \rightarrow \mathrm{g}}\left( t \right) \right|^{2} \approx \frac{\mathit{\Omega}_{0}^{2} \sin^{2} \left[ \left( \mathit{\Delta}_{\mathrm{eff}} -\omega \right) t/2 \right]}{4\left( \mathit{\Delta}_{\mathrm{eff}} -\omega \right)^{2}}.
\end{equation}

\label{B} \setcounter{equation}{0} \renewcommand{\theequation}{B%
\arabic{equation}}
\label{B} \setcounter{figure}{0} \renewcommand{\thefigure}{B%
\arabic{figure}}

\setcounter{section}{0} \renewcommand{\thesection}{APPENDIX B}
\section{Details of numerical simulations}
\label{appendix_b}

Here we present  the details of the numerical simulations of time evolution with time-dependent Hamiltonian $\hat{\mathcal{H}} \left( t \right)$.  The evolved state at time $t+\Delta t$ can be computed by the standard fourth-order Runge–Kutta method with the expansion \cite{carpenter1994fourth}
 \begin{equation}
 	\begin{aligned}| \Psi \left( t+\Delta t \right) \rangle =& | \Psi \left( t \right) \rangle +\frac{1}{6} | \Psi^{\left( 1 \right)} \rangle +\frac{1}{3} | \Psi^{\left( 2 \right)} \rangle \\
 	&+\frac{1}{3}| \Psi^{\left( 3 \right)} \rangle+\frac{1}{6} | \Psi^{\left( 4 \right)} \rangle +O(\Delta t^{5}),\end{aligned}
 \end{equation}
where four auxiliary vectors are 
\begin{equation}
	\begin{aligned}|\Psi^{\left( 1 \right)} \rangle &=-\mathrm{i} \hat{\mathcal{H}} \left( t \right) |\Psi \left( t \right) \rangle \Delta t\\ |\Psi^{\left( 2 \right)} \rangle &=-\mathrm{i} \hat{\mathcal{H}} \left( t+\frac{\Delta t}{2} \right) \left( |\Psi \left( t \right) \rangle +\frac{1}{2} |\Psi^{\left( 1 \right)} \rangle \right) \Delta t\\ |\Psi^{\left( 3 \right)} \rangle &=-\mathrm{i} \hat{\mathcal{H}} \left(  t+\frac{\Delta t}{2}\right) \left( |\Psi \left( t \right) \rangle +\frac{1}{2} |\Psi^{\left( 2 \right)} \rangle \right) \Delta t\\ |\Psi^{\left( 4 \right)} \rangle &=-\mathrm{i} \hat{\mathcal{H}} \left( t+ \Delta t \right) \left( |\Psi \left( t \right) \rangle +|\Psi^{\left( 3 \right)} \rangle \right) \Delta t\end{aligned}.
\end{equation}
The time steps of length is taken as $\Delta t=4\times 10^{-4}$ for the numerical computations.  In the implementation of this method, one can exploit the sparsity of the matrix representation of the Hamiltonian to save the computer memory and time.

\bibliography{Ref.bib}

\end{document}